\documentclass[aps,pra,reprint,longbibliography]{revtex4-1}
\usepackage{amsmath,amssymb,amsthm}
\usepackage{graphicx}
\usepackage{float}
\usepackage{mathtools}
\usepackage{braket}
\usepackage{qcircuit}
\usepackage{placeins}
\usepackage{enumitem}

\newtheorem{theorem}{Theorem}
\newtheorem{proposition}{Proposition}
\newtheorem{definition}{Definition}

\newcommand{\x}[1]{\text{#1}}
\newcommand{\rk}{\mathrm{rk}}
\newcommand{\im}{\mathrm{im}}
\newcommand{\pr}{\mathrm{pr}}
\newcommand{\poly}{\mathrm{poly}}
\newcommand{\R}{\mathcal{R}}
\newcommand{\idx}{\mathsf{idx}}
\newcommand{\cgate}[1]{\mathrm{c}\text{-}#1}
\newcommand{\ccgate}[1]{\mathrm{cc}\text{-}#1}
\newcommand{\notgate}{\mathsf{NOT}}
\newcommand{\andgate}{\mathsf{AND}}
\newcommand{\orgate}{\mathsf{OR}}
\newcommand{\xorgate}{\mathsf{XOR}}

\newcommand{\eq}[1]{\hyperref[eq:#1]{Eq.~(\ref*{eq:#1})}}
\newcommand{\fig}[1]{\hyperref[fig:#1]{Figure~\ref*{fig:#1}}}
\newcommand{\tab}[1]{\hyperref[tab:#1]{Table~\ref*{tab:#1}}}
\newcommand{\sect}[1]{\hyperref[sec:#1]{Section~\ref*{sec:#1}}}
\newcommand{\defn}[1]{\hyperref[def:#1]{Definition~\ref*{def:#1}}}
\newcommand{\prop}[1]{\hyperref[prop:#1]{Proposition~\ref*{prop:#1}}}
\newcommand{\thm}[1]{\hyperref[thm:#1]{Theorem~\ref*{thm:#1}}}

\usepackage[hidelinks]{hyperref}
\hypersetup{
    pdftitle={}, 
    pdfauthor={}, 
    colorlinks=true, 
    linkcolor=blue, 
    citecolor=blue, 
    urlcolor=blue, 
   	final=true, 
}

\begin{document}

\title{Possibilistic simulation of quantum circuits by classical circuits}

\author{Daochen Wang}\thanks{Department of Mathematics and Joint Center for Quantum Information and Computer Science, University of Maryland. Email: \texttt{wdaochen@gmail.com}}

\begin{abstract}
In breakthrough work, Bravyi, Gosset, and K\"{o}nig (BGK) [Science, 2018] unconditionally proved that constant-depth quantum circuits are more powerful than their classical counterparts. Their result is equivalent to saying that a particular family of constant depth quantum circuits takes classical circuits at least $\Omega(\log n)$ depth to ``simulate'', in a certain sense. In our paper, we formalise their sense of simulation, which we call ``possibilistic simulation'' or ``p-simulation'', and construct explicit classical circuits that can p-simulate any depth-$d$ quantum circuit with Clifford and $t$ $T$-gates in depth $O(d+t)$. Our classical circuits use $\{\notgate, \andgate, \orgate\}$ gates of fan-in~$\leq 2$.
\end{abstract}
\maketitle

\section{Introduction}

Quantum computation is widely believed to provide advantages over classical computation. Popular science articles sometimes explain the advantage by some notion of quantum parallelism. Indeed, it is true that a quantum computer can efficiently operate, ``in parallel'', upon a quantum wavefunction encompassing exponentially many classical states. Unfortunately, the class of efficient operations (standard quantum gates for example) is restrictive. Moreover, any quantum computation must finish with a measurement that collapses the quantum wavefunction to just one classical state. Even ignoring noise, these caveats mean it is not obvious if quantum computation holds any actual advantage.

Academically, belief in quantum advantage is more correctly supported by evidence of quantum-classical separations in query, time, and circuit complexity. 

In circuit complexity, one early result is Ref.~\cite{moore_parityfanout}, which showed that quantum circuits can compute in constant depth the parity of all input bits assuming the controlled-multi-$\notgate$ gate, $\cgate{X}^{\otimes{n}}$, can be implemented in constant depth (also see the later work, Ref.~\cite{hoyer_spalek_qc_fanout_powerful}). Separation is therefore \emph{provably} achieved because parity is provably uncomputable by constant depth classical circuits~\cite{arora_barak}. More precisely, the separation is against classical $\x{AC}^0$ circuits, where gates are restricted to $\{\notgate, \andgate, \orgate\}$ of arbitrary fan-in and fan-out and where circuit \emph{size}, i.e., the number of gates, is restricted to be polynomial~\footnote{We need to restrict the gate set, else a ``parity gate'' can compute parity in depth $1$. We need to restrict circuit size, else parity on $n$-bits can be computed in depth $3$, via the conjunctive normal form of parity (e.g., when $n=3$, $x_1\oplus x_2 \oplus x_3 = (x_1 \wedge \lnot x_2 \wedge \lnot x_3) \vee (\lnot x_1 \wedge  x_2 \wedge \lnot x_3) \vee (\lnot x_1 \wedge \lnot x_2 \wedge x_3) \vee (x_1 \wedge  x_2 \wedge  x_3)$), using an exponential number of gates in $\{\notgate, \andgate, \orgate\}$ of arbitrary fan-in and fan-out. The depth of $3$ comes from applying three layers of gates: a layer of $O(n2^n)$ $\notgate$ gates, followed by a layer of $O(2^n)$ $\andgate$ gates, followed by a single $\orgate$ gate.}. However, as the $\cgate{X}^{\otimes{n}}$ gate acts on all $n$ qubits, it is unreasonable to assume it can be implemented in constant depth.  Only recently, in breakthrough work by Bravyi, Gosset, and K\"{o}nig ~\cite{quantum_advantage_shallow_bravyi_gosset_konig} (henceforth BGK) was such unreasonable assumptions removed in achieving a circuit complexity separation. Indeed, their separation was achieved by a quantum circuit with gates in $\{H, \ccgate{Z}, \cgate{S^{\dagger}}\}$. What is particularly satisfying is that BGK proved their separation via Ref.~\cite{barrett_bgk} from quantum foundations, which can be viewed as extending fundamental Bell-type inequalities to a multi-party, bounded-locality setting. One can already catch a glimpse of the connection between circuits and foundations by noting that the BGK quantum circuit applies $\cgate{S^{\dagger}}$ gates followed by $H$ gates just before computational basis measurement. But this is the same as a controlled changing of measurement basis from $X$ to $Y$, a technique commonly used in optimal quantum strategies of non-local games like CHSH~\cite{chsh} or GHZ~\cite{ghz}.

Notwithstanding the build-up of evidence in favour of quantum advantage, substantial efforts have also been devoted to the time-efficient classical simulation of quantum computation. In this arena, the most celebrated result is arguably the Gottesman-Knill theorem which says that quantum Clifford circuits on $n$ qubits, whereby $\ket{0^n}$ is evolved by $L$ Clifford gates, i.e., $\{H, S, \cgate{X}\}$~\footnote{For concreteness in our paper, ``Clifford gates'' always means $\{H, S, \cgate{X}\}$ gates. None of our results would essentially change if we say ``Clifford gates'' are one-qubit and two-qubit gates \emph{generated} by $\{H, S, \cgate{X}\}$. Our results do change if we say ``Clifford gates'' are arbitrary multi-qubit gates generated by $\{H, S, \cgate{X}\}$. This change is unimportant unless such gates also have constant depth physical implementations.} and followed by $M$ Pauli-observable measurements, can be efficiently simulated in time $O((L+M)n^{3})$~\cite{gottesman_phd,nielsen_chuang,stabiliser_aaronson_gottesman}. 

One main motivation for studying simulation is to understand quantum advantage better. For example, the Gottesman-Knill theorem means that entanglement is insufficient for time-complexity quantum advantage because Clifford circuits can generate entanglement~\cite{entanglement_stabilizer_fattal_cubitt}. 

Currently, there are two well-established notions of simulating a given quantum circuit~\cite{vandennest_simulation,jozsa_simulation,pashayan_simulation}: strong and weak. Strong simulators approximate the probability of a particular output, while weak simulators approximately sample from the output distribution. 

In our paper, we extract from recent Refs.~\cite{quantum_advantage_shallow_bravyi_gosset_konig,coudron_vidick_stark_2018,average_case_le_gall,schaeffer_ac0, bgkt_noise,grier_nc1} another notion of simulation. We say a (non-uniform) classical circuit simulates a quantum circuit if, \emph{over all inputs}, the output of the classical circuit is a \emph{possible} (i.e., occurring with non-zero probability) output of the quantum circuit. We call this ``possibilistic simulation'' or ``p-simulation''. Then, BGK's result can be phrased as an unconditional $\Omega(\log n)$ lower bound on (even non-uniform) classical circuits that p-simulate certain constant depth quantum circuits. 

It is known that p-simulating (classically controlled) Clifford circuits, like those appearing in BGK, is in the complexity class $\oplus \textsf{L}\subset\textsf{NC}^2$~\cite{stabiliser_aaronson_gottesman,grier_nc1}. This means that there exists an $O(\log^2 n)$-depth uniform classical circuit that p-simulates the BGK quantum circuits. 

In comparison, our main result is the construction of non-uniform classical circuits that can p-simulate \emph{any} depth-$d$ quantum circuit with Clifford and $t$ $T$-gates in depth $O(d+t)$ (\thm{clifford_T}). We consider Clifford and $T$-gates as they are universal for quantum computation. 

\section{Possibilistic simulation}

In this section, we give our formal definition of p-simulation, as extracted from Refs.~\cite{quantum_advantage_shallow_bravyi_gosset_konig, coudron_vidick_stark_2018, average_case_le_gall,schaeffer_ac0,bgkt_noise}. 

\begin{definition}\label{def:setup}
We make the following definitions for circuits with $n$ input lines and $m$ output lines.

\begin{itemize}
    \item A relation on the Cartesian product $\{0,1\}^{n}\times \{0,1\}^{m}$ is a subset $\R\subseteq \{0,1\}^{n}\times \{0,1\}^{m}$.
    
    \item A quantum circuit $Q$ on $n$ input qubit lines and measured on $m$ output qubit lines in the computational basis defines a relation $\R(Q)\subseteq \{0,1\}^{n}\times \{0,1\}^{m}$ by:
    \begin{equation}\label{eq:nonzero_condition}
        (x,y)\in \R(Q) \iff \Bra{y}Q\Ket{x} \neq 0.
    \end{equation}
    
    \item Let $C: \{0,1\}^n \rightarrow \{0,1\}^m$ be a classical circuit, and $\R$ be a relation on $\{0,1\}^{n}\times \{0,1\}^{m}$. We say $C$ p-simulates $\R$ if:
    \begin{equation}\label{eq:simulate_relation}
    (x,C(x))\in \R, \text{ for all } x\in \{0,1\}^{n}.
    \end{equation}
\end{itemize}
\end{definition}

In our paper, we follow BGK in restricting our classical circuits to having gates in the standard set $\{\notgate, \andgate, \orgate\}$ ($\{\lnot, \wedge, \vee\}$) of fan-in~$\leq 2$ but arbitrary fan-out. A gate's fan-in (fan-out) is its number of input (output) lines. The cost of our simulator stated in \thm{clifford_T} does require the gates to have arbitrary fan-out, as we briefly explain following \thm{clifford_T}. Also following BGK, we allow quantum circuits to use additional all-zero ``advice'' bitstring inputs.

\begin{definition}\label{def:possibilistic_sim}
Let $Q$ and $C$ be quantum and classical circuits, respectively. We say $C$ p-simulates $Q$ if $C$ p-simulates $\R(Q)$.
\end{definition}

For example, we can set $m=n=1$, and verify that $C = 0$ and $C=\x{NOT}$ p-simulates $Q=H$ (Hadamard gate) and $Q=X$ (Pauli X gate) respectively.

We note that the above definition of p-simulation can be generalized to the bounded-error and average-case setting -- see \sect{conclusion}.

\section{p-simulator construction}

In this section, we construct a p-simulator by explicitly constructing its classical circuit. We then analyse the cost of this p-simulator in terms of its circuit depth and size to prove the main result of this paper, \thm{clifford_T}.

We assume for simplicity that $m=n$ and that the quantum circuit takes no advice. It is simple to generalize this construction when these conditions do not hold.

We first construct classical circuits that p-simulate Clifford circuits and then extend to Clifford+$T$ circuits. The correctness of our constructions should be self-evident.

\textit{Clifford.} Let $Q$ be a Clifford circuit. First, we can write an $n$-bit input $\ket{x} = \ket{x_1\dots x_n}$ as $\ket{x}=X_{1}^{x_1} \cdots X_{n}^{x_n}\ket{0^{n}}$.

\begin{table*}
\begin{centering}
\begin{tabular}{l|l|l}
    $HX = ZH$ \ & 
    \ $HY = -YH$ \ & 
    \ $HZ = XH$, \\
    
    $SX = YS$ \ & 
    \ $SY = -XS$ \ & 
    \ $SZ = ZS$, \\
    
    $EX_1 = X_1X_2 E$ \ & 
    \ $EY_1 = Y_1X_2 E$ \ & 
    \ $EZ_1 = Z_1 E$,  \\

    $EX_2 = X_2E$ \ & 
    \ $EY_2 = Z_1 Y_2 E$ \ & 
    \ $EZ_2 = Z_1Z_2 E$. \\
\end{tabular}
\par\end{centering}
\caption{{\scriptsize{}Elementary commutation relations. For tidiness, we write $E$ for $\cgate{X}_{2}$ in this table only. The same commutation relations hold (up to global minus signs irrelevant for p-simulation) when there is the same exponent $e \in \{0, 1\}$ on the Pauli operator of the left-hand side and the Pauli operator(s) of the right-hand side. For example, the top left equation gives $HX^{e}=Z^{e}H$ for $e\in \{0,1\}$}.}
\label{tab:commutation_table}
\end{table*}

Now, we may use the commutation relations listed in \tab{commutation_table} to commute all $X_{i}^{x_{i}}$ past the Clifford circuit $Q$ and just before (computational basis) measurements. Note that $Q$ would remain unchanged. Moreover, we may wlog (without-loss-of-generality) assume that the resulting $x$-dependent gates on qubit $i\in[n]$ are of the form $X_{i}^{a^{(i)} \cdot x}$ for some $a^{(i)}\in\{0,1\}^{n}$, where the dot means inner product mod $2$ . The ``wlog'' is with respect to our definition of p-simulation because just before (computational basis) measurements, $Y$ can be replaced by $X$, and $Z$ by identity. Note that the $X$ gate is the same as the \emph{classical} $\notgate$ gate and we use the latter notation in the following.

Now, to p-simulate $Q$, simply pre-compute an $n$-bit string $s$ in the support of $Q\ket{0^{n}}$, which can be done efficiently by Gottesman-Knill. It is important to note that the pre-computation only helps construct the classical circuit which we first-and-foremost want to show \emph{exists}, so, in principle, it does not matter if pre-computation is inefficient as will be the case later. Then $s$ defines a classical circuit $C$ which, on input $x\in\{0,1\}^{n}$, outputs the $n$-bit string:
\begin{equation}\label{eq:clifford_classical}
    C(x) \coloneqq \left(\prod_{i=1}^{n}\notgate_{i}^{a^{(i)}\cdot x} \right) \, s = (\notgate^{s_i}(a^{(i)}\cdot x))_{i=1}^n.
\end{equation}

Writing $|\cdot|$ for the Hamming weight, it is clear that $a^{(i)} \cdot x$ can be computed in parallel, across $i\in[n]$, in depth $O(\log \max_{i} |a^{(i)}|)$ by an $\xorgate$-binary-tree of size $O(\sum_{i=1}^n |a^{(i)}|) = O(n\, \max_{i} |a^{(i)}|)$. $\xorgate$ can be replaced by its optimal decomposition into $4$ standard gates, i.e., ${\xorgate}(x,y) = (x\vee y)\wedge\lnot(x\wedge y)$. $s$ can be incorporated in depth $1$ via at most $n$ $\notgate$ gates. Therefore, $C$ can have depth $O(\log \max_{i} |a^{(i)}|)$ and size $O(n\, \max_{i} |a^{(i)}|)$. This completes the description of our construction in the Clifford case.

\begin{figure}[ht]
\includegraphics[angle=270,scale=0.25]{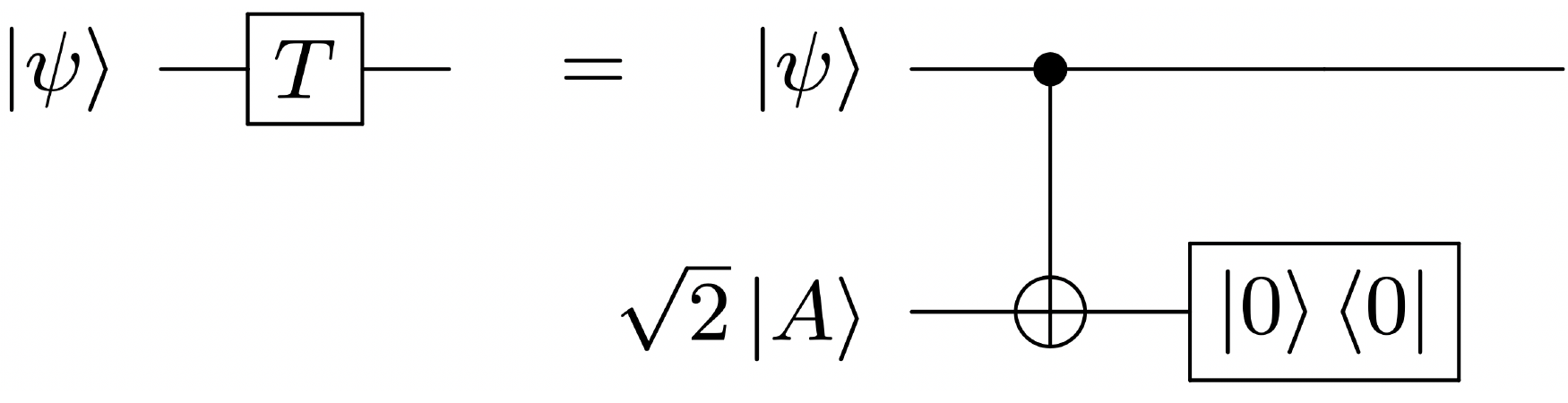}
\caption{{\scriptsize{}The $T$-gadget post-selected on $\ket{0}$. $\ket{A}$ is the so-called magic state $\frac{1}{\sqrt{2}}(\ket{0}+e^{i\pi/4}\ket{1})$. $\ket{0}\bra{0}$ is the post-selection projector onto $\ket{0}$ and can be performed just before measuring the original qubit. If we had post-selected on $\ket{1}$, we would implement $T^\dagger$ instead (up to a global phase).}}
\label{fig:T-gadget}
\end{figure}

\textit{Clifford+$T$.} Let $\tilde{Q}$ be a quantum circuit with Clifford gates and $t$ $T$-gates. We may replace each $T$-gate by a (post-selected) $T$-gadget, as shown in \fig{T-gadget}. Such replacement gives a \textit{Clifford} circuit $Q$ on $n+t$ qubits.

$Q$ has original input $\ket{x}$ on the top $n$ qubit lines and magic state inputs $\ket{A^{\otimes t}}$ on the bottom $t$ qubit lines. Just before measurements of the top $n$ qubit lines, $Q$ is post-selected for $\ket{0^{t}}$ in the bottom $t$ qubit lines. This construction is standard~\cite{bravyi_gosset_simulation}.

As in the Clifford case, we again write $\ket{x} = \ket{x_1\dots x_n}$ as $\ket{x}=X_{1}^{x_1} \cdots X_{n}^{x_n}\ket{0^{n}}$ and commute all $X_{i}^{x_{i}}$ past the Clifford circuit $Q$. This results (again wlog) in $Q$ followed by $X_{i}^{a^{(i)} \cdot x}$ on qubit $i\in[n+t]$, for some $a^{(i)}\in\{0,1\}^{n}$.

Next, we pre-compute the state $\ket{\psi}\coloneqq Q\ket{0^{n}}\ket{A^{\otimes t}}$. Note that this pre-computation is inefficient in general and is the reason why our circuit construction is non-uniform. In contrast to the Clifford case, it is believed that this pre-computation cannot be done efficiently, as else we can efficiently strongly simulate quantum computation. From $\ket{\psi}$, we pre-compute the $2^{t}$ states $\ket{\psi_{z}} \coloneqq \left(\mathbb{I}^{n} \otimes \bra{z}\right)\ket{\psi}$ where $z\in\{0,1\}^t$. $\ket{\psi_{z}}$ are necessarily non-zero $n$-qubit states equal to the output of $\tilde{Q}$ but with a $z$-indicated subset of $T$-gates replaced by  $T^{\dagger}$. Let $s(z)$ be an $n$-bit string in the support of $\ket{\psi_{z}}$. $s(z)$ defines a classical circuit $C_{z}$ which, on input $x\in\{0,1\}^{n}$, outputs the $n$-bit string:
\begin{equation}\label{eq:ai_1_to_n}
    C_{z}(x) \coloneqq \left(\prod_{i=1}^{n}\notgate_{i}^{a^{(i)}\cdot x}\right) \ s(z) = (\notgate^{s(z)_i}_i(a^{(i)}\cdot x))_{i=1}^n,
\end{equation}
where $a^{(i)} \cdot x$ can again be computed in depth $O(\log \max_{i} |a^{(i)}|)$. Up to this point, we have only used the $T$-gadget and commutation to define quantities. 

\begin{figure}
\includegraphics[angle=270,scale=0.275]{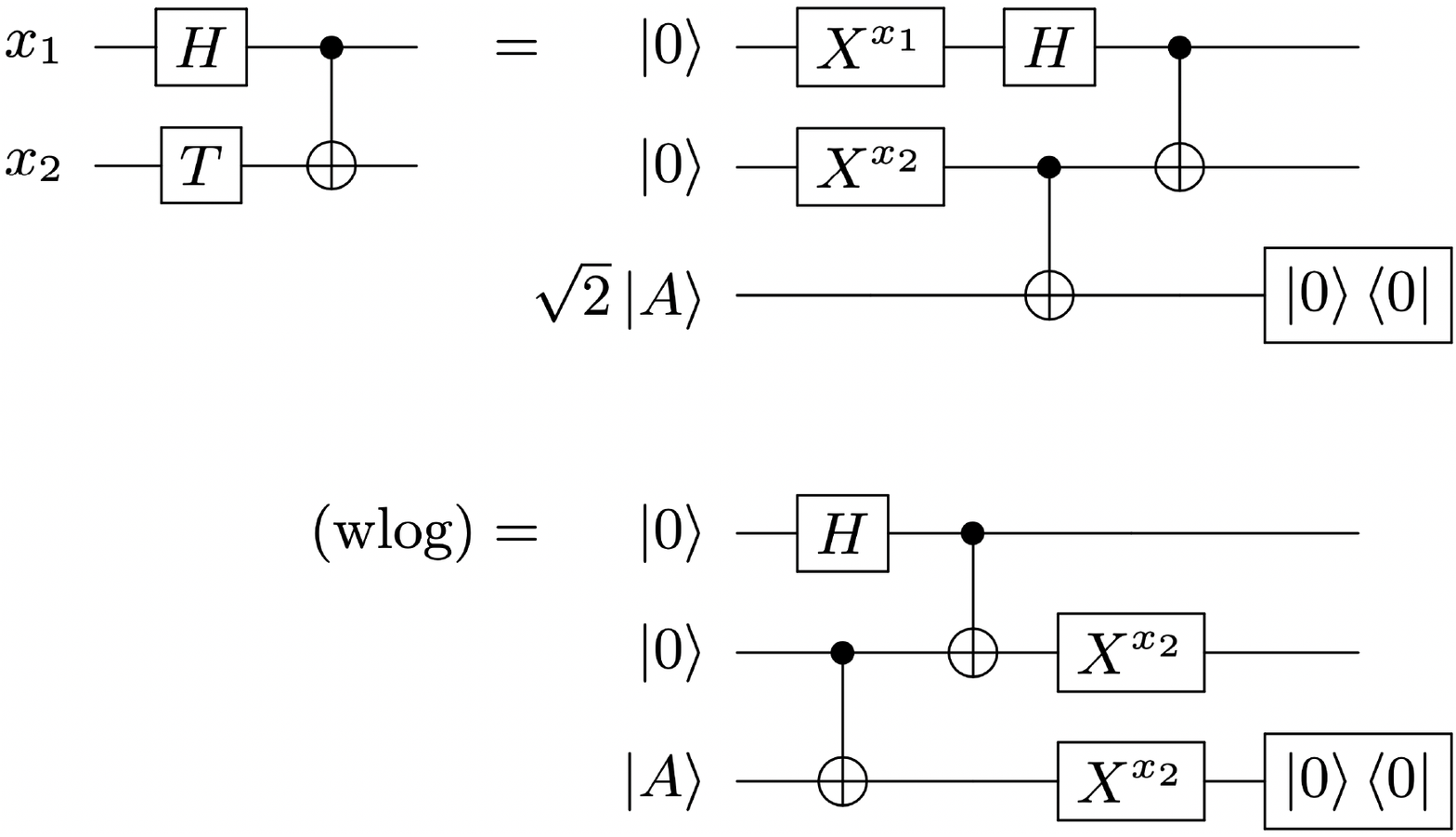}
\caption{\scriptsize{}Quantum circuit identities used to define quantities in our construction as illustrated by an example with $n=2$, $t=1$. Note that the removal of the global $\sqrt{2}$ factor is also wlog.}
\label{fig:b_example}
\end{figure}

In \fig{b_example}, we give an example with $n=2$, $t=1$, and where the quantities defined are (or can be):
\begin{align}
    & a^{(1)} = 000, \ a^{(2)} =  a^{(3)}= 010, \\
    &\ket{\psi} = \frac{1}{2}(\ket{000}+\ket{110} + e^{i\pi/4}\ket{001} + e^{i\pi/4} \ket{111}), \\
    & \ket{\psi_{0}}, \  \ket{\psi_{1}} \propto  \ket{00}+\ket{11},\\
    & s(0) = 00, \ s(1) = 11.
\end{align}

In order to describe the classical p-simulation circuit $C$, let $A \in \mathbb{F}_2^{t\times n}$ denote the $t \times n$ matrix with entries $A_{ij} = a^{(n+i)}_{j}\in \mathbb{F}_2 = \{0,1\}$ for all $i\in[t],j\in[n]$. Let $\rk(A)$ and $\im(A)$ denote the rank and image of $A$ respectively. Note that $|\im(A)| = 2^{\rk(A)}$.

We proceed to describe $C$. $C$ takes as input $x\in\{0,1\}^{n}$ and consists of three consecutive stages.

In Stage 1, we compute the $2^{\rk(A)}$ $n$-bit strings $C_{z}(x)$, for all $z \in \im(A)$, in depth $O(\log \max_{i} |a^{(i)}|)$ using $O(\sum_{i=1}^n |a^{(i)}| + n 2^{\rk(A)}) = O(n (\max_{i} |a^{(i)}| + 2^{\rk(A)}))$ gates. In the gate count, the term $\sum_{i=1}^n |a^{(i)}|$ is due to computing the string $c\coloneqq (a^{(i)}\cdot x)_{i=1}^n$ and the term $n2^{\rk(A)}$ is due to applying up to $n$ $\notgate$ gates, more precisely $\{\notgate_i^{s(z)_i}\}_{i=1}^n$, to $c$ for each $z\in \im(A)$, cf. \eq{ai_1_to_n}.

In Stage 2, we compute the $t$-bit string:
\begin{equation}\label{eq:ai_n_1_to_n_t}
    z(x) \coloneqq \left(\prod_{i=n+1}^{n+t}\notgate_{i}^{a^{(i)}\cdot x}\right) \, 0^t = (a^{(n+i)}\cdot x)_{i=1}^t,
\end{equation}
in depth $O(\log \max_{i} |a^{(i)}|)$ using $O(\sum_{i=1}^t |a^{(n+i)}|) = O(t \max_{i} |a^{(i)}|)$ gates. Note that $z(x) = Ax \in \im(A)$.

In Stage 3, we implement a simple switching circuit~\footnote{In v1 of this paper, step 2 of Stage 3 was disregarded and not costed which led to major errors.}. (This construction may be better understood after first examining the proof of \prop{general_circuit}.) More specifically, we compute the $n$-bit output string $y\coloneqq C_{z(x)}(x)$ in two serial steps:
\begin{enumerate}[label=(\roman*)]
\item Compute a $2^{\rk(A)}$-bit string $f$ of Hamming weight $1$, where $f_{j} = \delta_{a(j),z(x)}$ and $a(j)$ is the $(j+1)$-th $t$-bit string in $\im(A)$ (under any fixed enumeration) for $j\in \{0,\dots,2^{\rk(A)}-1\}$, in depth $O(\log t)$ using $O(t 2^{\rk(A)})$ gates via the formula:
\begin{equation}
    f_j = \bigwedge_{k=1}^t \big[\notgate^{a(j)_k \oplus 1} \, z(x)_k\big],
\end{equation}
where we used the fact that $\delta_{u,v} = \notgate^{u\oplus 1} \, v$ for any two bits $u,v\in \{0,1\}$.

\item Compute the $n$-bit string $y$ in depth $O(\rk(A))$ using $O(n 2^{\rk(A)})$ gates via the formula:
\begin{equation}
y_{i} = \bigvee_{j=0}^{2^{\rk(A)}-1} \Big[[C_{a(j)}(x)]_{i} \wedge f_{j}\Big].
\end{equation}
\end{enumerate}

\begin{figure*}
\includegraphics[scale=0.70]{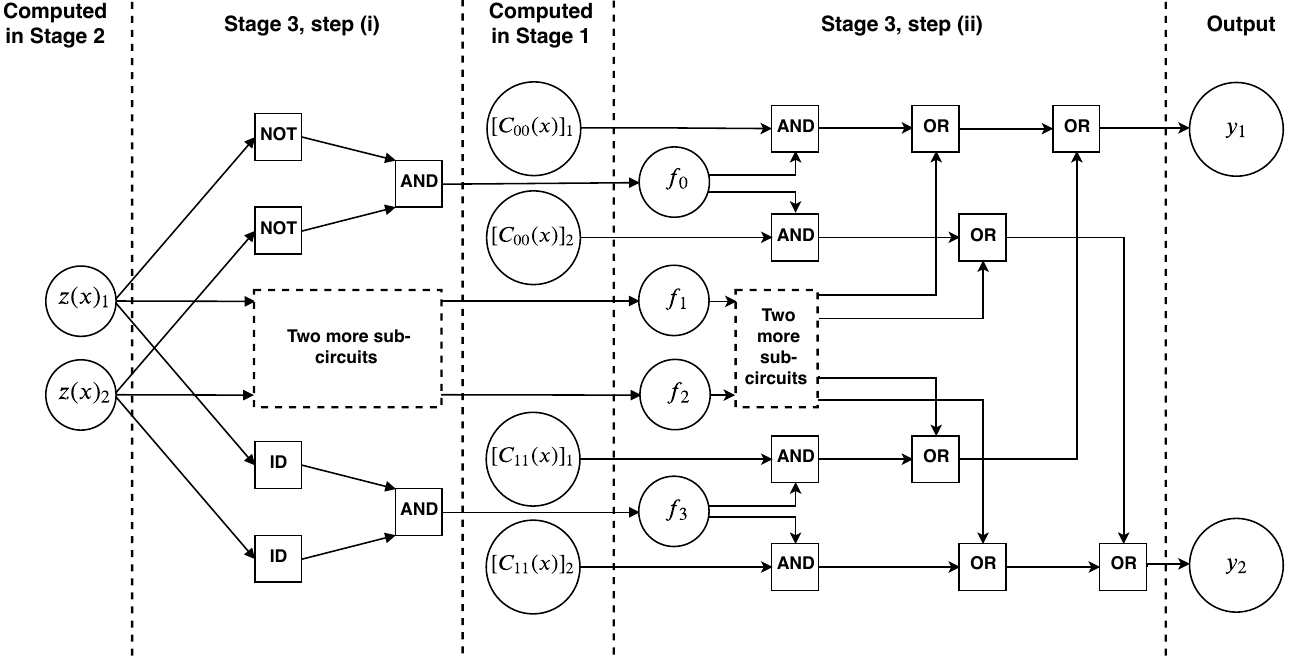}
\centering{}\caption{{\scriptsize{}Illustration of our construction with $n=t=\rk(A)=2$ and input $x$, showing how Stages~1--3 fit together in series. Circles are single bits and squares are gates. The notations $z(x), C_{z}(x)$, and $f_{j}$ are defined in \eq{ai_n_1_to_n_t}, \eq{ai_1_to_n}, and the description of Stage~3 respectively. $z(x)$, $C_{z}(x)$ are $(t=2)$-bit and $(n=2)$-bit strings respectively, on which a subscript $i$ denotes the $i$-th bit. Note that each gate has fan-in~$\leq2$.}}
\label{g plot}
\end{figure*}

We illustrate our overall circuit in the case $n=t=\rk(A)=2$ in Fig.~\ref{g plot}. This completes the description of our construction. 

We now analyse the circuit depth and size of our construction to obtain the main result of this paper.

\begin{theorem}\label{thm:clifford_T}
Any $n$-qubit quantum circuit $Q$ of depth $d$ with Clifford, $t$ $T$-gates, and associated $A$ matrix can be $p$-simulated by a classical circuit $C$ of:
\begin{alignat*}{3}
    &\text{depth} &&= O(d+\log(t) +\rk(A)) &&= O(d+t),
    \\
    &\text{size} &&= O((n+t)
\, (2^{\rk(A)}+n)) &&= O((n+t)(2^t+n)),
\end{alignat*}
that consists of $\{\notgate,\andgate,\orgate\}$ gates of fan-in $\leq 2$ and arbitrary fan-out.
\end{theorem}

\begin{proof}
Define $C$ by our construction applied to $Q$. The depth and size of $C$ can be analysed as follows.

In \eq{ai_1_to_n} and \eq{ai_n_1_to_n_t}, we have: 
\begin{equation}
    |a^{(i)}| = \min(O(2^{d}),n), \quad \text{for all} \  i\in[n+t],
\end{equation}
because $Q$ has depth $d$ with Clifford gates of fan-in~$\leq2$, and the Hamming of weight of $a^{(i)}\in \{0,1\}^n$ is at most $n$. So Stages 1 and 2 can be implemented by a circuit of depth $O(d)$ and size $O(n(n+t+2^{\rk(A)}))$. As discussed, Stage 3 can be implemented by a circuit of depth $O(\log(t) + \rk(A))$ and size
$O((n+t)2^{\rk(A)})$. Now, note that $\rk(A)\leq t$ because $A$ is a $t\times n$ matrix. Therefore, $C$ has overall depth $O(d+\log(t) + \rk(A)) = O(d+t)$ and overall size
\begin{equation}
\begin{aligned}
&\, O((n+t)n + n2^{\rk(A)} + (n+t)2^{\rk(A)})
\\
=& \, O((n+t)
\, (2^{\rk(A)}+n))
\\
=& \, O((n+t)(2^t+n)),
\end{aligned}
\end{equation}
as required.
\end{proof}

The cost of our p-simulator in \thm{clifford_T} does require the gates to have arbitrary fan-out as assumed in its statement. However, it can be seen that if the fan-out is bounded by a constant, then the theorem still holds but with an additional depth of
\begin{equation}\label{eq:additional_depth}
    O(\rk(A)+\log(n) + \log(t)),
\end{equation} 
and an additional size of 
\begin{equation}\label{eq:additional_size}
    O((n+t) (2^{\rk(A)} + n)).
\end{equation}

These additional costs are due to additional gates used to fan out (i.e., copy) variables at each of the three stages of our construction. The details are as follows.
\begin{enumerate}[wide, labelwidth=1pt, labelindent=0pt]
    \item[\emph{Stage 1.}] For each $i\in [n]$, we use $O(\log n)$ depth and $O(n)$ gates to make $n$ copies of input variable $x_i$. Similarly, for each $i \in [n]$, we use $O(\log(2^{\rk(A)})) = O(\rk(A))$ depth and $O(2^{\rk(A)})$ gates to make $2^{\rk(A)}$ copies of $a^{(i)} \cdot x$. Therefore, over all $i\in [n]$, these copying steps of Stage 1 cost a depth of $O(\rk(A)+\log n)$ and size of $O(n (2^{\rk(A)} + n))$.
    \item[\emph{Stage 2.}] For each $i\in [n]$, we use $O(\log t)$ depth and $O(t)$ gates to make $t$ copies of input variable $x_i$. Therefore, over all $i\in [n]$, this copying step of Stage 2 costs a depth of $O(\log t)$ and size of $O(nt)$.
    \item[\emph{Stage 3.}] Step (i). For each $k\in [t]$, we use $O(\log 2^{\rk(A)}) = O(\rk(A))$ depth and $O(2^{\rk(A)})$ gates to make $2^{\rk(A)}$ copies of $z(x)_k$. Step (ii). For each $j \in \{0,\dots,2^{\rk(A)}-1\}$, we use $O(\log n)$ depth and $O(n)$ gates to make $n$ copies of $f_j$. Therefore, over all $k\in [t]$ and $j\in \{0,\dots,2^{\rk(A)}-1\}$, these copying steps of Stage 3 cost a depth of $O(\rk(A) + \log n)$ and size of $O(t2^{\rk(A)} + n2^{\rk(A)})$.
\end{enumerate}
Adding together the additional depths and additional sizes in each of the three stages gives \eq{additional_depth} and \eq{additional_size}, respectively.

\paragraph*{Additional p-simulation techniques.} We can also refine and extend \thm{clifford_T} by thinking more carefully about our construction. First, we may choose $s(z)$ more carefully such that the size of the set $\{s(z) \mid  z \in \im(A) \}$ is minimised. Second, Pauli-$T$ commutation relations, namely $TZ=ZT$, $TX \propto(X+Y)T$, and $TY\propto(X-Y)T$, instead of the $T$-gadget, sometimes suffice to handle a $T$-gate, which removes its constant depth contribution. Third, the only property of the $T$-gate that is used is that it can be applied by state injection into a Clifford circuit. Since this property holds for any gate that is in the set $\mathcal{G}$ consisting of diagonal gates~\cite{bravyi_browne_calpin_etal} and gates in the third-level of the Clifford hierarchy~\cite{gottesman_chuang_injection}, \thm{clifford_T} extends to circuits composed of $\mathcal{G}$ gates and Clifford gates, i.e., $t$ could count the number of gates the circuit has in $\mathcal{G}$ (that are not Clifford) and the theorem would still hold. Since an arbitrary single-qubit gate can be decomposed into three $Z$-rotation gates (which are diagonal) and two Hadamard gates~\cite[Problem 8.1]{ksv_book}, \thm{clifford_T} also extends to circuits composed of single-qubit gates and Clifford gates. In fact, since an arbitrary constant-qubit gate can be decomposed into a constant number of single-qubit gates and CNOT gates~\cite[Section 4.5.2]{nielsen_chuang}, \thm{clifford_T} also extends to circuits composed of constant-qubit gates and Clifford gates.

\section{Comparison to strong and weak simulators}

There are two pre-existing notions of simulation, strong simulation and weak simulation. Following the notation of \defn{setup} with $m=n$, a strong simulation of a quantum circuit $Q$ (with a fixed input) is an (approximate) \emph{evaluation} of the probability of obtaining a given output bitstring $y\in \{0,1\}^n$ when $Q$ is measured in the computational basis at the end of the computation. A weak simulation of $Q$ is an (approximate) \emph{sample} of $y \in \{0,1\}^n$ from the distribution arising from measuring $Q$ in the computational basis at the end of the computation. A strong \emph{simulator} of a family $\mathcal{F}$ of quantum circuits is a classical algorithm that takes as input a (classical description of) quantum circuit $Q\in \mathcal{F}$ and $y\in \{0,1\}^n$ and performs strong simulation of $(Q,y)$. A weak simulator of $\mathcal{F}$ is a classical algorithm that takes as input a circuit $Q\in \mathcal{F}$ and performs weak simulation of $Q$. For more details about these definitions, see, e.g., Refs.~\cite[Section 2]{vandennest_simulation}, \cite[Section 2]{jozsa_simulation}, or \cite[Section 8.1]{pashayan_simulation}. For a recent review of strong and weak simulators, see Ref.~\cite[Section III]{pashayan_fast}. 

We now argue that existing results on strong and weak simulators do not imply our result on p-simulation, \thm{clifford_T}. Our argument is also intended to elucidate the differences between p-simulation and strong and weak simulation.

We first consider using a strong simulator for p-simulation. We claim that even if it costed \emph{zero} depth and size for the strong simulator to evaluate the probability $\pr(y)$ of measuring $y$ for each $y\in \{0,1\}^n$, it would still cost depth $\Omega(n)$ and size $\Omega(2^{n})$ for the strong simulator to output a $y$ that has $\pr(y)\neq 0$. We show the claim by the following argument. For a quantum circuit $Q$, we  define the family of $2^n$ quantum circuits $\mathcal{F}_Q \coloneqq \{Q_x \mid x\in \{0,1\}^{n}\}$, where $Q_x$ is $Q$ but with the input $x$ hardwired at the beginning of $Q$ using Pauli $X$ gates. To use a strong simulator $\mathcal{A}$ to $p$-simulate $Q$, we should apply it to the family $\mathcal{F}_Q$ and each $y\in \{0,1\}^n$. Now, we set $Q$ to be a quantum circuit that for an input $x$ outputs $x$ with probability $1$ ($Q$ doesn't necessarily have to be the identity circuit and could be complicated). For a given $Q_x$, we may wlog assume that we have used $\mathcal{A}$ to compute $\pr(y)$ for all $y\in \{0,1\}^n$, since we assumed this costs zero depth and size. Then, the computation remaining is to output $x\in \{0,1\}^n$ given a $2^n$-bit string $(\pr(0^n),\dots, \pr(1^n))$, where $\pr(x) = 1$ and $\pr(y) = 0$ for all $y\in \{0,1\}^n$ with $y\neq x$. In other words, the computation remaining is the computation of the function $\idx:\{0,1\}^{[2^n]}\rightarrow \{0,1\}^n$, where the input $z$ is promised to have Hamming weight $1$, and the output $f(z)$ equals the $i\in [2^n]$ such that $z_i = 1$. Our initial claim then follows from:
\begin{proposition}
Let $\idx$ be defined as above. Then, any classical circuit $C$ with fan-in $\leq 2$ that computes $\idx$ must have depth $\Omega(n)$ and size $\Omega(2^n)$.
\end{proposition}

\begin{proof}
We first establish the size lower bound. Consider the $2^n/2$ input bit pairs $(z_1,z_2),(z_3,z_4),\dots, (z_{2^n-1},z_{2^n})$. We claim that within each pair there must exist at least one bit that is the input to a gate in $C$. Suppose for contradiction that neither $z_i$ nor $z_{i+1}$ is input to a gate, then the output of $C$ is independent of $z_i$ and $z_{i+1}$. Therefore, the outputs of $C$ on inputs $z^{(i)}$ and $z^{(i+1)}$ are the same, where $z^{(j)}$ denotes the $2^n$-bit string of Hamming weight $1$ with exactly one $1$ at position $j$. This is a contradiction since $\idx(z^{(i)})=i\neq i+1 = \idx(z^{(i+1)})$ and $C$ computes $\idx$. Hence the claim. Therefore, there are at least $2^n/2$ inputs to gates in $C$. But each gate in $C$ takes at most $2$ inputs by the fan-in condition. Therefore, $C$ must have at least $2^n/4$ gates. 

Now, we establish the depth lower bound.  Suppose $C$ has depth $d$, then $C$ has at most $O(n2^d)$ gates, where the $n$ arises from $C$ having $n$-bit output and the $2^d$ arises from $C$ having fan-in $\leq 2$. Therefore $c \cdot n2^d \geq 2^n/4$ for some constant $c$. Hence $d\geq\Omega(n)$.
\end{proof}

Therefore, using a strong simulator for p-simulation is worse than using our classical p-simulator except when $d+t = \Omega(n)$ (see \thm{clifford_T}).

We note that some strong simulators have the extra ability to evaluate certain marginal probabilities of the output $y$, meaning that they can evaluate the probability that certain subsets of bits of $y$ take given values. These strong simulators can be used as weak simulators \cite[Lemma 1]{jozsa_simulation}. Also see Ref.~\cite{bravyi_marginals} for another approach for reducing weak to strong simulation that does not involve evaluating marginal probabilities.

We now consider using a weak simulator for p-simulation.  Observe that an \emph{exact} weak simulator for the circuit family $\mathcal{F}_Q$ defined above is a p-simulator for $Q$ since a sample output by the exact weak simulator must occur with non-zero probability. Therefore, $p$-simulation is strictly easier than exact weak simulation.

The weak simulators most comparable to our p-simulator are those in Refs.~\cite{bravyi_gosset_simulation, bravyi_browne_calpin_etal, seddon_quantifying} that weakly simulate Clifford+$T$ circuits by exploiting stabilizer decompositions. Indeed,  our construction is inspired by Ref.~\cite{bravyi_gosset_simulation}. However, we note the weak simulators in Refs.~\cite{bravyi_gosset_simulation, bravyi_browne_calpin_etal, seddon_quantifying} only sample from a distribution that is $\epsilon$-close to the output distribution of $Q$ and run in time $\Omega(1/\epsilon^r)$ for some $r>0$. As $\epsilon$ cannot be set to zero, these weak simulators are necessarily non-exact and so are incomparable to our p-simulator: they might output a sample that is output by $Q$ with zero probability, something our p-simulator never does.

Aside from the issue of exact sampling, another issue is that weak simulators are typically costed in terms of time complexity rather than circuit depth or size complexity. The time complexity of the weak simulators in Refs.~\cite{bravyi_gosset_simulation, bravyi_browne_calpin_etal, seddon_quantifying} take the form $O(\poly(n,g) + \poly(t) 2^{\beta t})$, where $n$ is the number of qubits, $g$ is the number of one- or two-qubit Clifford gates, $t$ is the number of $T$-gates, and $0<\beta < 1$. Since a computation taking time $\mathcal{T}$ can be implemented by a circuit of size $O(\mathcal{T}\log(\mathcal{T}))$~\cite[Proof of Theorem 6.6]{arora_barak}, this means that these weak simulators can be implemented using circuits of size $\tilde{O}(\poly(n,g) + \poly(t) 2^{\beta t})$, where the tilde hides logarithmic factors. The dominant term is $2^{\beta t}$ which is better than the dominant term in the size cost of our p-simulator, i.e., $2^t$, since $\beta<1$. In addition, these circuits have the advantage of being efficiently computable~\cite[Remark 6.7]{arora_barak}, which is not the case for our p-simulator circuit.

However, the size of a circuit says little about its depth. Indeed, it is not obvious how to deduce the depth bound in \thm{clifford_T} even with $t=0$ by considering a Gottesman-Knill simulator, i.e., an exact weak simulator that operates according to the proof of the Gottesman-Knill theorem in Refs.~\cite{nielsen_chuang,stabiliser_aaronson_gottesman}. While a Gottesman-Knill simulator can update each of $n$ stabilisers in parallel, updating the sign of each after, say, a Hadamard layer $H^{\otimes{n}}$, uses depth $O(\log n)$. Worse still, measurement in the standard basis, i.e.,  measurement of $n$ Pauli observables $Z_{i}$ for $i\in[n]$, uses sequential depth $O(n)$ and does not seem easily parallelisable. This issue is addressed with some work in Ref.~\cite[Appendix~C of arXiv version]{grier_nc1}, where the authors show that exact weak simulation of (even classically controlled) Clifford circuits of any depth is in $\oplus \textsf{L}\subset\textsf{NC}^2$, and so can be implemented by classical circuits of depth $O(\log^2 n)$. Nevertheless, it is still unclear how to recover the depth bound in \thm{clifford_T} by considering weak simulators when $t>0$.

When $t\geq n/\beta$, the weak simulators in Refs.~\cite{bravyi_gosset_simulation, bravyi_browne_calpin_etal, seddon_quantifying} become essentially trivial because a weak simulator that operates simply by storing and updating the quantum state as a length-$2^n$ vector has a comparable time complexity of $O((g+t)2^n)$~\cite{aaronson_complexity}. 

A similar phenomenon occurs with our p-simulator. When $t \geq n$, the depth and size of our p-simulator (as stated in \thm{clifford_T}) become essentially trivial because \emph{any} function $f: \{0,1\}^n\rightarrow \{0,1\}^n$ can be computed by a simple circuit of comparable depth and size. The last fact can be seen by considering a circuit similar to Stage 3 of our p-simulator. For completeness, we prove it below.

\begin{proposition}\label{prop:general_circuit}
Let $f: \{0,1\}^n \rightarrow \{0,1\}^n$ be an arbitrary function. Then, there is a classical circuit $C$ with fan-in $\leq 2$ of depth $O(n)$ and size $O(n2^n)$ that computes $f$.
\end{proposition}

\begin{proof}
Let $x\in \{0,1\}^n$ be the input to $f$. The circuit $C$ computes $f(x)$ in two serial steps. In the first step, $C$ computes the $2^n$ bits $\{f_z \mid z \in \{0,1\}^n\}$, defined by $f_z = 1$ if and only if $z=x$, using the formula
\begin{equation}
    f_z = \bigwedge_{i=1}^n (\notgate^{z_i\oplus 1} \, x_i),
\end{equation}
where we used the fact that $\delta_{u,v} = \notgate^{u\oplus 1} \, v$ for any two bits $u,v\in \{0,1\}$. This takes depth $O(\log n)$ and size $O(n 2^n)$. In the second step, $C$ computes the output $y\in \{0,1\}^n$ by the formula
\begin{equation}
    y_i = \bigvee_{z\in \{0,1\}^n} (f(z))_i \wedge f_z.
\end{equation}
This takes depth $O(n)$ and size $(n2^n)$. 

Adding together the depths and sizes in the first and second steps gives the result.
\end{proof} 

Unfortunately, the above observation means our p-simulator gives a trivial result if applied to p-simulate the BGK quantum circuit~\cite[Fig.~1 of arXiv version]{quantum_advantage_shallow_bravyi_gosset_konig}. Indeed, in the BGK quantum circuit, there are $\Omega(n^2)$ $\ccgate{Z}$ gates and $\Omega(n)$ $\cgate{S}$ gates.  If we decompose each $\ccgate{Z}$ and $\cgate{S}$ gate into a constant number of $T$ and Clifford gates, then we obtain a p-simulator of the BGK quantum circuits with depth $\Omega(n^2)$. This is very inefficient because, as we noted at the beginning, there exists a p-simulator of depth $O(\log^2 n)$.

\section{Conclusion}\label{sec:conclusion}
In p-simulation, we have defined a natural framework that precisely captures the new type of quantum advantage that has recently come to light~\cite{quantum_advantage_shallow_bravyi_gosset_konig, coudron_vidick_stark_2018, average_case_le_gall,schaeffer_ac0,bgkt_noise}. We found how $T$ gates are necessary for advantage according to \thm{clifford_T}. In particular, we find that Clifford quantum circuits do not yield quantum advantage and that BGK's use of (classically) \textit{controlled}-Clifford gates is vital. More generally, our paper helps motivate and preclude new candidate quantum circuits that exhibit advantage. 

Our work raises at least two interesting questions:
\begin{enumerate}
    \item Can a p-simulator of Clifford+$T$ circuits have depth scaling as $t^\alpha$ and size scaling as $2^{\alpha t}$ for some constant $0<\alpha<1$, where $t$ is the number of $T$ gates? One approach may be to consider the (approximate) stabilizer decomposition~\cite{bravyi_smith_smolin, bravyi_gosset_simulation, bravyi_browne_calpin_etal} of the state $\ket{A^{\otimes t}}$. This approach has improved the time complexity of strong and weak simulators from scaling with $2^t$ to scaling with $2^{\beta t}$ for some constant $0< \beta < 1$.
    \item Can we reduce the depth and size of our p-simulator if we generalize the definition of p-simulation to the bounded-error and average-case setting? In this setting, we generalize the condition in \eq{simulate_relation} to
    \begin{equation}
        \pr((x,C(x))\in \R)\geq 1-\delta,
    \end{equation}
    where $\delta\in(0,1)$ and the probability is over some probability distribution over the input $x$ and the randomness in the classical circuit $C$. In particular, it would be interesting to see if there exist more efficient p-simulators (under the generalized definition) when $x$ is distributed according to the hard probability distributions described in, for example, Refs.~\cite{quantum_advantage_shallow_bravyi_gosset_konig,coudron_vidick_stark_2018,average_case_le_gall,schaeffer_ac0}.
\end{enumerate}

\section{Acknowledgements}
\label{sec:acknowledgements}
I thank Luke Schaeffer for finding a critical error in my costing the total depth of Stage 3 of the construction in v1 as $O(\log t)$ (cf.~footnote~[29]), sharing an early draft of Ref.~\cite{schaeffer_ac0}, and useful discussions. I thank Matt Coudron, David Gosset, Tongyang Li, Carl Miller, and Aarthi Sundaram for useful discussions. I also thank anonymous reviewers for their helpful comments, suggestions, and corrections.

I acknowledge support from the Army Research Office (grant W911NF-20-1-0015); the Department of Energy, Office of Science, Office of Advanced Scientific Computing Research, Accelerated Research in Quantum Computing program; and the National Science Foundation (grant DMR-1747426).

\bibliography{references}
\end{document}